\documentclass[11 pt,runningheads]{llncs}
\usepackage{graphicx}
\usepackage{makecell}

\usepackage{array}
\usepackage{hyperref}
\hypersetup{linkcolor=blue,colorlinks=true}
\hypersetup{urlcolor=blue, colorlinks=true}
\hypersetup{citecolor=blue, colorlinks=true}
\usepackage{caption}
\renewcommand{\figurename}{Figure}
\renewcommand{\tablename}{Table}
\makeatletter
\newcommand{\@chapapp}{\relax}%
\usepackage[toc,page]{appendix}
\usepackage{IEEEtrantools}

\usepackage{circuitikz}
\usepackage{tikz}
\usetikzlibrary{arrows,shapes.gates.logic.US,shapes.gates.logic.IEC,calc}
\usetikzlibrary{shapes.geometric, arrows}
\usetikzlibrary{shapes,arrows,fit,calc,positioning,automata}

\usepackage{mathtools}
\usepackage{xparse}
\usepackage{csquotes}


\usepackage{amssymb}
\usepackage{xr}
\usepackage{multirow}% http://ctan.org/pkg/multirow
\usepackage{hhline}% http://ctan.org/pkg/hhline
\usepackage{fixltx2e}
\usepackage{commath}
\usepackage{algorithm}
\usepackage{algpseudocode}
\usepackage{amsmath}
\algdef{SE}[DOWHILE]{Do}{doWhile}{\algorithmicdo}[1]{\algorithmicwhile\ #1}%

\begin{document}
\title{Security Analysis of WG-7 Lightweight Stream Cipher against Cube Attack}
\author{Bijoy Das \inst{1}\and Abhijit Das\inst{2} \and Dipanwita Roy Chowdhury\inst{3}}
\authorrunning{B. Das et al.}
\institute{Indian Institute of Technology Kharagpur\\
\email{mantunsec@gmail.com}
\and Indian Institute of Technology Kharagpur\\
\email{abhij@cse.iitkgp.ac.in}
\and Indian Institute of Technology Kharagpur\\
\email{drc@cse.iitkgp.ac.in}}
\maketitle              % typeset the header of the contribution
%

%%%% 6. ABSTRACT %%%%
\begin{abstract}

Welch--Gong (WG) is a hardware-oriented LFSR-based stream cipher. WG-7 is a version of
the eStream submission Welch--Gong, used for RFID encryption and authentication purposes. It offers 80-bit cryptographic security.
In modern days, almost all ciphers achieve the security by exploiting the nonlinear feedback structure. In this paper, we investigate the security of the nonlinear feedback-based initialization phase of the WG-7 stream cipher using the conventional bit-based division property of cube attack, by considering
the cipher in a non-blackbox polynomial setting. In our work, we mount the cube attack using mixed-integer-linear-programming
(MILP) models. The results
of our attack enable us to recover the secret key of WG-7 after 20 rounds of
initialization utilizing $2^{10}$ keystream bits in $2^{73}$ time. We show that our
proposed attack takes significantly lower data complexity. To the best of our knowledge, our attack is the first one that investigates the security of the nonlinear feedback-based initialization phase of WG-7 cipher.


\end{abstract}

%%%% 7. PAPER CONTENT %%%%

\section{Introduction}\label{sec:intro}%

WG-7~\cite{LuoCGL10} is a fast lightweight word-oriented stream cipher whose construction
is based on the Welch-Gong~(WG)~\cite{NawazG08} stream cipher. WG-7
includes a 23-stage word-oriented LFSR with each stage working over the finite
field $\mathbb{F}_7$ , and a nonlinear filter function that is based on the WG transformation.
First, the LFSR is loaded with the key and the IV. Next, the LFSR with
its nonlinear function is run for 46 iterations. Then the
cipher generates the appropriate keystream bits that are used for encryption. The cipher is mainly designed for encryption in resource-constrained environments such as
mobile phones, smart cards, and RFID applications.

We see that 80-bit lightweight stream cipher WG-7 has already been broken with an algebraic attack.
The authors of \cite{OrumiehchihaPS12,DBLP:journals/dcc/Ronjom17} mounted the attack separately on WG-7 by exploiting
the fact that WG-7 is updated linearly in the KSG
phase. In~\cite{OrumiehchihaPS12}, it is possible for the attacker to find the annihilator function $g$ of the
filtering function $f$ such that $fg=0$ and $\deg g<\deg f=5$. The best annihilator
function $g$ with degree 3 is worked out. This helps the attacker to mount the
algebraic attack on WG-7 with time complexity about $2^{27}$ and data complexity
$2^{19.38}$. In case of~\cite{DBLP:journals/dcc/Ronjom17}, the authors improved the algebraic attack complexity of~\cite{OrumiehchihaPS12} by providing the upper bound for the \textit{spectral immunity} (SI) of the cipher. But the computation of SI depends on the number of nonzero coefficients in the polynomial. The measurement of nonzero coefficients is bounded by the degree of the polynomial. But the degree of any polynomial does not change when the feedback path contains only linear operation. This helps the attacker to improve the attack with $2^{17.3}$ data complexity and $2^{28}$ time complexity.
But this cannot happen in presence of the nonlinear function in the feedback path of the cipher, where the degree of the functions grows very fast.

So we observe that the aforementioned attacks work only when the feedback-path of the cipher is linear. In this paper, we analyze the initialization phase which contains the nonlinear operation in the feedback path, and show that the secret key of the WG-7 cipher can be recovered to the reduced 20 rounds of the initialization phase using only $2^{10}$ data complexity. To the best of our knowledge, our attack is the first one to investigate the security of the nonlinear feedback-based initialization phase of the WG-7 lightweight stream cipher.

\noindent
\text{\bf Our Contributions. } We mount the cube attack on the initialization phase of the reduced 20 rounds of WG-7. We construct the model of the division trail that propagates through the
WG-permutation (WGP) function of WG-7 as a 7-bit S-box trail. Moreover, we build the model of the division trails through the linear layer of the cipher by finding the invertible (sub)matrices of the matrix of the linear transformation. Our optimizations lead to a full key recovery using only $2^{10}$ bits
in the keystream and with a time complexity of $2^{73}$. Table~\ref{table:comp_} shows the comparison with the existing algebraic attack. We see that our attack is not impaired by the nonlinear function in the feedback
operation. Our attack also significantly reduces data complexity of $2^{10}$.

\begin{table}[!h]
\begin{center}
\caption{Performance on Data Complexity}
\label{table:comp_}
\scalebox{1}{
\begin{tabular}{|c|c|c|c|} 
\hline
 Ref. & \thead{Data \\Complexity} & \thead{Time \\ Complexity} & \thead{Work Environment}\\
\hline
\cite{OrumiehchihaPS12} &  $2^{19.38}$ &$2^{27}$ & \text{It works only in the absence}\\
\cline{1-3}
~\cite{DBLP:journals/dcc/Ronjom17} & $2^{17.3}$ & $2^{28}$& \text{of the nonlinear feedback path}\\
\hline
\thead{Our\\Approach}& ${\bf 2^{10}}$ &$2^{73}$ & \thead{\bf Attack become successfull even if\\ \bf nonlinear function is \\ \bf present in the feedback path}\\

\hline

\end{tabular}
}
\end{center}
\end{table}

\noindent
The rest of the paper is organized as follows. In Section~\ref{sec:prilim}, we review
the concepts of cube attack. Section~\ref{sec:DP_BDP}
presents a brief overview of division property, bit-based division property,
and how to model the division trails using Mixed Integer Linear Programming (MILP). The specification
of WG-7 is provided in Section~\ref{sec:Desc}. Section~\ref{sec:cube_wg7} elaborates
our proposed cube attack on the initialization phase of WG-7. Section~\ref{sec:conclusion} concludes the paper.

\section{Preliminaries}\label{sec:prilim}%
Here, we give the notations and definitions we will use in this paper. 
\subsection{\textbf{Cube Attack}}\label{subsec:cubeAttack}%
The cube attack was proposed by Dinur and Shamir in EUROCRYPT~\cite{dinur2009cube} to recover secret key. For an $n_1$-bit key $k=(k_1,k_2,\ldots,k_{n_1})$ and $m_1$-bit IV $v=(v_1,v_2,\ldots,v_{m_1})$, let $f(x)$ be a boolean function from $\mathbb{F}^n_2$ to $\mathbb{F}^1_2$ such that $x = k||v$ and $n=n_1+m_1$. Let $\bf u\in\mathbb{F}^n_2$ be a constant vector. Then the ANF of \textit{f}($x$) is defined as $f(x)=x^u\times p(x) + q(x)$, where each term of $q(x)$ is not divisible by $x^u$. For a set of \textit{cube indices} $I = \{0\leq i \leq n-1 : u_i=1\}\subset \{0,1,\ldots,n-1\}$, $x^u$ represents the corresponding monomial.
Therefore, the summation of \textit{f}($x$) over all values of $C_{I} = \{x\in \mathbb{F}^n_2 : u \succeq x\}$ is given by
\begin{equation}
%\Large
\label{eqn:superpoly}
    \bigoplus\limits_{x\in C_I}^{} f(k,v) = \bigoplus\limits_{x\in C_I}^{} (x^u\times p(x)
  + q(x)) = p(x).
\end{equation}

\noindent
where $p(x)$ is called the \textit{superpoly} of $C_I$, and it only involves the variables $x_j$ such that $u_j=0$ for $0\leq j \leq n-1$.

Equation~(\ref{eqn:superpoly}) implies that if the attacker gets a superpoly that
is simple enough, she can query the encryption oracle feeding $C_I$. All the
first keystream bits returned are summed to evaluate the right-hand side of
Equation~(\ref{eqn:superpoly}). Subsequently, she recovers the secret key bits
by solving a system of equations.

\subsection{MILP-Aided Bit-Based Division Property}\label{sec:DP_BDP}%

{\bf Bit Product Function} $\pi_u(x)$. For any $u\in \mathbb{F}^{n}_2$, let $\pi_u(x)$ be a function from $\mathbb{F}^{n}_2$ to $\mathbb{F}_2$. For any input $x\in \mathbb{F}^{n}_2$, $\pi_u(x)$ is the AND of $x[i]$ satisfying $u[i]=1$. It is defined as
 $\pi_u(x) = \displaystyle\prod_{i=1}^{n}x[i]^{u[i]}$

\begin{definition}[\textit{Division Property}]
 Let $X$ be a multi-set whose elements take values from
$\mathbb{F}^{l_0}_2 \times \mathbb{F}^{l_1}_2 \times \cdots \times \mathbb{F}^{l_{m-1}}_2$.
The multi-set $X$ has the division property $D^{l_0,l_1,\ldots,l_m-1}_\mathbb{K}$,
where $\mathbb{K}$ denotes a set of $m$-dimensional vectors whose $i$-th elements take values
between 0 and $l_i$, if it fulfills the following condition:
 \[
    \bigoplus_{x\in\mathbb{X}}\pi_u(x)= 
\begin{cases}
    \textit{unknown}, \text{ if there exist } k\in\mathbb{K} \text{ such that } \textit{wt(u)}\succeq k,\\
    0,  \hspace{3 cm}  \text{otherwise.}
\end{cases}
\]
\end{definition}

\noindent
If there are $k, k^{'}\in\mathbb{K}$ such that $k\succeq k^{'}$
in the division property $D^{l_0,l_1,\ldots,l_{m-1}}_\mathbb{K}$, then $k$ can be removed from $\mathbb{K}$ because it is redundant. 
When $l_0,l_1,\ldots,l_{m-1}$ are restricted to 1, we talk about
\textit{bit-based division property}.
The main idea of MILP-aided bit-based division property is to model the
propagation rules as a series of linear (in)equalities. We adopt the
MILP models for \textbf{copy}, \textbf{AND} and \textbf{XOR} from~\cite{TodoIHM17}.
\section{The WG-7 Stream Cipher}\label{sec:Desc}%
\begin{figure}[!htb]
  \centering
  \includegraphics[height = 3.5cm, width=\linewidth]{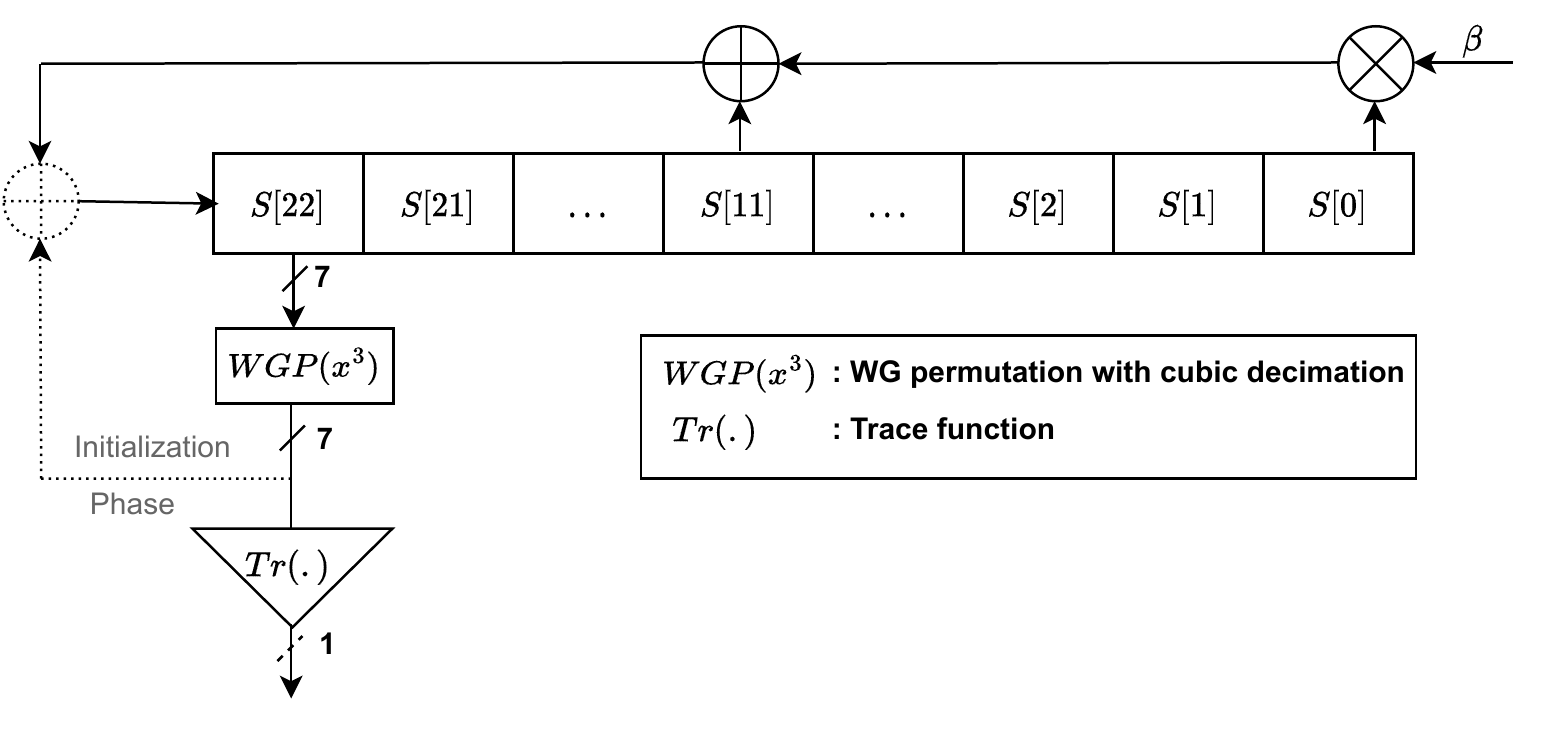}
  \caption{Design of WG-7 Stream Cipher}
  \label{fig:WG7_design}
\end{figure}

WG-7~\cite{LuoCGL10} is a stream cipher designed by Y.~Luo, Q.~Chai, G.~Gong
and X.~Lai in 2010. As depicted in Figure~\ref{fig:WG7_design}, the structure of
WG-7 consists of a 23-stage LFSR and a nonlinear filtering
function which is realized by the Welch--Gong (WG) transformation. Each stage
of WG-7 works over the finite field $\mathbb{F}_{2^7}$. This finite field is defined by
the primitive polynomial $g(x)=x^7+x+1$. The characteristic polynomial is primitive over
$\mathbb{F}_{2^7}$, and is given by 
 $f(x) = x^{23}+x^{11}+\beta \in\mathbb{F}_{2^{7}}[x]$ \text{,  } where $\beta$ is a root of $g(x)$.
 
\noindent
This cipher uses an 80-bit secret key and an 81-bit initialization vector (IV). The cipher
works in two phases: initialization and Keystream Generation (KSG) phase. In the initialization phase, the cipher is first loaded with the 80-bit key and the 81-bit IV. Then, the LFSR is clocked for 46 times with the output of a nonlinear permutation feedback function WGP, whereas the KSG phase does not include this nonlinear feedback path.
We denote the state at the $i$-th round by
$S^i=S^i[0]||S^i[1]||\ldots||S^i[22]$, where
$S^i[j]=(s^i_{7j},s^i_{7j+1},s^i_{7j+2},s^i_{7j+3},s^i_{7j+4},s^i_{7j+5},s^i_{7j+6})$
for $0\leq j\leq 22$.
The $n$-bit secret key and the $m$-bit IV are represented as $K_{0,\ldots,{n-1}}$ and $IV_{0,\ldots,{m-1}}$,
respectively. Initially, the 23-stage LFSR is loaded as follows: For $0\leq i\leq 10$, $S^{0}[2i]=(K_{7i,7i+1,7i+2,7i+3},IV_{7i,7i+1,7i+2})$, $S^{0}[2i+1]=(K_{7i+4,7i+5,7i+6},IV_{7i+3,7i+4,7i+5,7i+6})$
and $S^{0}{[22]}=(K_{77,78,79},IV_{77,78,79,80})$.
The state-update function in this phase is given by
$S^{i+1}[j]=S^{i}[j+1]$ for $0\leq j \leq 21$,
and $S^{i+1}[22]=S^{i}[11]\oplus \beta S^{i}[0]  \oplus WGP(S^{i}[22])$.

During the KSG phase, the keystream bit
is given by $z_{i-46} = Tr(s^3 + s^9 + s^{21} + s^{57} + s^{87})$, where $s = S^i[22] =
(s^i_{154},s^i_{155},s^i_{156},s^i_{157},s^i_{158},s^i_{159},s^i_{160})$. The ANF
representation of the keystream bit is represented by
\noindent
$z_{i-46} = s^i_{160}+s^i_{158}s^i_{160} + s^i_{158}s^i_{159} + s^i_{157}+s^i_{157}s^i_{160} + s^i_{157}s^i_{159} + s^i_{157}s^i_{159}s^i_{160}+s^i_{157}s^i_{158}+s^i_{157}s^i_{158}s^i_{160}+s^i_{157}s^i_{158}s^i_{159}s^i_{160}+s^i_{156}s^i_{158}s^i_{160}+s^i_{156}s^i_{157}+s^i_{156}s^i_{157}s^i_{159}s^i_{160}+s^i_{156}s^i_{157}s^i_{158}s^i_{159}s^i_{160}+s^i_{155}+s^i_{155}s^i_{159}+s^i_{155}s^i_{159}s^i_{160}+ s^i_{155}s^i_{156}s^i_{158} +s^i_{154}s^i_{155}+s^i_{155}s^i_{156}s^i_{160} +s^i_{155}s^i_{158}s^i_{159}s^i_{160}+s^i_{155}s^i_{157}s^i_{159}s^i_{160}+s^i_{155}s^i_{157}s^i_{158}s^i_{159}s^i_{160}+s^i_{155}s^i_{156}s^i_{157}s^i_{160}+s^i_{155}s^i_{156}s^i_{157}s^i_{158}+s^i_{154}+s^i_{154}s^i_{159}+s^i_{154}s^i_{159}s^i_{160}+s^i_{154}s^i_{156}+s^i_{154}s^i_{158}s^i_{159}+s^i_{154}s^i_{157}s^i_{160}+s^i_{154}s^i_{157}s^i_{159}s^i_{160}+s^i_{154}s^i_{157}s^i_{158}s^i_{159}s^i_{160}+\\s^i_{154}s^i_{156}s^i_{160}+s^i_{154}s^i_{156}s^i_{158}s^i_{160}+s^i_{154}s^i_{156}s^i_{158}s^i_{159}s^i_{160}+s^i_{154}s^i_{156}s^i_{157}s^i_{159}+s^i_{154}s^i_{156}s^i_{157}s^i_{158}s^i_{159}+s^i_{154}s^i_{155}s^i_{159}s^i_{160}+s^i_{154}s^i_{155}s^i_{158}s^i_{160}+s^i_{154}s^i_{155}s^i_{156}+s^i_{154}s^i_{155}s^i_{158}s^i_{159}s^i_{160}+s^i_{154}s^i_{155}s^i_{157}s^i_{159}+s^i_{154}s^i_{155}s^i_{157}s^i_{158}s^i_{160}+\\s^i_{154}s^i_{155}s^i_{157}s^i_{158}s^i_{159}+s^i_{154}s^i_{155}s^i_{156}s^i_{159}+s^i_{154}s^i_{155}s^i_{156}s^i_{159}s^i_{160}+\\s^i_{154}s^i_{155}s^i_{156}s^i_{158}s^i_{159}+s^i_{154}s^i_{155}s^i_{156}s^i_{157}s^i_{160}+s^i_{154}s^i_{155}s^i_{156}s^i_{157}s^i_{158}$.

\noindent
In this phase, the state is updated as 
$S^{i+1}[j]=S^{i}[j+1]$ for $0\leq j \leq 21$, \text{ and }
$S^{i+1}[22]= S^{i}[11]\oplus \beta S^{i}[0]$.

\begin{algorithm}[thb]
 \caption{MILP model for $\beta.X$ operation in WG-7}
\label{algo:milp_Linear_Layer}
\begin{algorithmic}[1]

  \Function{LinearLayer}{X}

\State Let $M$ be the $7\times 7$ Linear Transformation Matrix for WG-7 Cipher
\State $M^{'} \gets$ MatrixTranspose(M)\Comment{Compute the Transpose of M}
   \State $\mathcal{M}.var \gets y_{j} \text{ as binary for }0\leq j \leq 6$
\For {$i \in \{0,1,2,3,4,5,6\}$}

\State $\mathcal{M}.var\gets t_{ij}$ \text{ as binary} { if }$M_{i,j}=1$ \text{ for } $0\leq j \leq 6$ 
\State $\mathcal{M}.var\gets t_{ji}$ \text{ as binary} { if }$M^{'}_{i,j}=1$ \text{ for } $0\leq j \leq 6$

\EndFor

\For {$i \in \{0,1,2,3,4,5,6\}$}
\State $\mathcal{M}.con \gets x_i = \sum{}^{}[t_{ji} \text{ if }  M^{'}_{i,j}=1 \text{ for } 0\leq j \leq 6] \text{ where }X=(x_0,x_1,\ldots,x_6)$
\State $\mathcal{M}.con \gets \sum{}^{}[t_{ij} \text{ if }  M_{i,j}=1 \text{ for } 0\leq j \leq 6] = y_i$

\EndFor

\State $\mathcal{M}.con \gets \sum\limits_{j=0}^{6}y_{j} = \sum\limits_{j=0}^{6}x_{j}$

  \State {\Return {($\mathcal{M},[y_0,y_1,y_2,y_3,y_4,y_5,y_6]$})}
  
  \EndFunction

\end{algorithmic}
\end{algorithm}

\section{Cube Attack on WG-7}\label{sec:cube_wg7}%
In this section, we describe about the process of mounting the cube attack on WG-7 cipher.
\subsection{\bf Model the Initialization Phase of WG-7 Using MILP}\label{subsec:milp}%
We start the cube attack by modeling the division property propagation in each round
for each of the functions used in the WG-7 cipher.

\begin{algorithm}[!htb]
 \caption{MILP model for KSG operation in WG-7}
\label{algo:milp_KSG}
\begin{algorithmic}[1]

  \Function{KSG}{S}
   \State $\mathcal{M}.var \gets z \text{ as binary }$
   \State \text($\mathcal{M}$, \text{A}, $a_1)\gets$ \text{AND}(S,[158,160]) \text{,}
   \text($\mathcal{M}$, \text{B}, $a_{2})\gets$ \text{AND}(\text{A},[154,155,157,158,160])

   \State \text($\mathcal{M}$, \text{C}, $a_3)\gets$ \text{AND}(\text{B},[158,159]) \text{,}
   \phantom{00}\text($\mathcal{M}$, \text{D}, $a_{4})\gets$ \text{AND}(\text{C},[154,155,157,159])

   \State \text($\mathcal{M}$, \text{E}, $a_5)\gets$ \text{AND}(\text{D},[157,160]) \text{,}
   \phantom{001}\text($\mathcal{M}$, \text{F}, $a_{6})\gets$ \text{AND}(\text{E},[154,155,158,160])

   \State \text($\mathcal{M}$, \text{G}, $a_7)\gets$ \text{AND}(\text{F},[157,159])\text{,}
   \phantom{000}\text($\mathcal{M}$, \text{H}, $a_{8})\gets$ \text{AND}(\text{G},[154,155,159,160])

   \State \text($\mathcal{M}$,\text{I}, $a_9)\gets$ \text{AND}(\text{H},[157,159,160])\text{,}
   \phantom{00000}\text($\mathcal{M}$, \text{J}, $a_{10})\gets$ \text{AND}(\text{I},[154,156,160])

   \State \text($\mathcal{M}$, \text{K}, $a_{11})\gets$ \text{AND}(\text{J},[157,158]) \text{,}
   \phantom{0}\text($\mathcal{M}$, \text{L}, $a_{12})\gets$ \text{AND}(\text{K},[156,157,159,160])

   \State \text($\mathcal{M}$, \text{M}, $a_{13})\gets$ \text{AND}(\text{L},[157,158,160])\text{,}
   \phantom{0}\text($\mathcal{M}$, \text{N}, $a_{14})\gets$ \text{AND}(\text{M},[156,158,160])

   \State \text($\mathcal{M}$, \text{O}, $a_{15})\gets$ \text{AND}(\text{N},[157:160]) \text{,}
   \text($\mathcal{M}$, \text{P}, $a_{16})\gets$ \text{AND}(\text{O},[154,156,158,160])

   \State \text($\mathcal{M}$,\text{Q}, $a_{17})\gets$ \text{AND}(\text{P},[156,157]) \text{,}
   \phantom{0}\text($\mathcal{M}$, \text{R}, $a_{18})\gets$ \text{AND}(\text{Q},[155,158,159,160])

   \State \text($\mathcal{M}$, \text{T}, $a_{19})\gets$ \text{AND}(\text{R},[156:160])\text{,} 
    \phantom{00001}\text($\mathcal{M}$, \text{U}, $a_{20})\gets$ \text{AND}(\text{T},[154,157,160])

   \State \text($\mathcal{M}$, \text{V}, $a_{21})\gets$ \text{AND}(\text{U},[155,159])\text{,}
   \text($\mathcal{M}$, \text{W}, $a_{22})\gets$ \text{AND}(\text{V},[155,157,159,160])

   \State \text($\mathcal{M}$,\text{X}, $a_{23})\gets$ \text{AND}(\text{W},[155,159,160])\text{,}
   \phantom{0}\text($\mathcal{M}$, \text{Y}, $a_{24})\gets$ \text{AND}(\text{X},[155,156,160])

   \State \text($\mathcal{M}$, \text{Z}, $a_{25})\gets$ \text{AND}(\text{Y},[155,[157:160]])\text{,}
   \phantom{0}\text($\mathcal{M}$, \text{b}, $a_{26})\gets$ \text{AND}(\text{Z},[155,156,158])

   \State \text($\mathcal{M}$, \text{c}, $a_{27})\gets$ \text{AND}(\text{b},[155:158])\text{,}
   \phantom{001}\text($\mathcal{M}$, \text{d}, $a_{28})\gets$ \text{AND}(\text{c},[154,157,159,160])

   \State \text($\mathcal{M}$, \text{e}, $a_{29})\gets$ \text{AND}(\text{d},[154,159])\text{,}
   \phantom{000}\text($\mathcal{M}$, \text{f}, $a_{30})\gets$ \text{AND}(\text{e},[154,156,157,159])
   
   \State \text($\mathcal{M}$, \text{g}, $a_{31})\gets$ \text{AND}(\text{f},[154,156])\text{,}
   \phantom{01}\text($\mathcal{M}$, \text{h}, $a_{32})\gets$ \text{AND}(\text{g},[154,156,[158:160]])

   \State \text($\mathcal{M}$, \text{k}, $a_{33})\gets$ \text{AND}(\text{h},[[155:157],160])\text{,}
   \phantom{00}\text($\mathcal{M}$, \text{l}, $a_{34})\gets$ \text{AND}(\text{k},[154,159,160])

   \State \text($\mathcal{M}$,\text{m}, $a_{35})\gets$ \text{AND}(\text{l},[154,158,159])\text{,}
   \phantom{00}\text($\mathcal{M}$,\text{q}, $a_{36})\gets$ \text{AND}(\text{m},[154,[157:160]])

   \State \text($\mathcal{M}$, \text{r}, $a_{37})\gets$ \text{AND}(\text{q},[154,155])\text{,}
   \phantom{00}\text($\mathcal{M}$, \text{t}, $a_{38})\gets$ \text{AND}(\text{r},[154,155,[158:160]])

   \State \text($\mathcal{M}$, \text{u}, $a_{39})\gets$ \text{AND}(\text{t},[154,155,[157:159]])\text{,}
   \phantom{00}\text($\mathcal{M}$,\text{v}, $a_{40})\gets$ \text{AND}(\text{u},[154:156])

   \State \text($\mathcal{M}$,\text{w}, $a_{41})\gets$ \text{AND}(\text{v},[154,[156:159]])\text{,}
   \text($\mathcal{M}$, \text{x}, $a_{42})\gets$ \text{AND}(\text{w},[[154:156],159])
   
   \State \text($\mathcal{M}$, \text{y}, $a_{43})\gets$ \text{AND}(\text{x},[[154:156],159,160])\text{,}
   \phantom{00}\text($\mathcal{M}$, \text{z}, $a_{44})\gets$ \text{AND}(\text{y},[154:158])
   
   \State \text($\mathcal{M}$, \text{i}, $a_{45})\gets$ \text{AND}(\text{z},[[154:156],158,159])
   \text($\mathcal{M}$, \text{j}, $a_{46})\gets$ \text{AND}(\text{i},[[154:157],160])

   \State $\text(\mathcal{M}, S_{47}, a_{47})\gets \text{XOR}(\text{j},[154,155,157,160])$
   \State $\mathcal{M}.con \gets z = \sum_{1}^{47}a_i$  
  \State {\Return {($\mathcal{M},S_{47},z$})}
  \EndFunction

\end{algorithmic}
\end{algorithm}

\noindent
\textbf{MILP model for the Feedback Function (FBK)}. The function FBK at the $i$-th
round is expressed as $\beta S^{i}[0] \oplus S^{i}[11]$.
To model $\beta S^{i}[0]$, one can use its ANF representation described in
Section~\ref{sec:Desc}. Let $(x_0,x_1,\ldots,x_6)$
and $(y_0,y_1,\ldots,y_6)$ be the input and output of $\beta S^{i}[0]$, respectively.  
 Using the technique~\cite{sun2019milp}, we get the following system of equations by introducing 17 intermediate binary variables $t_i$ for $1\leq i \leq 17$ (see equation~(\ref{eq:example_left_right2})).

 \begin{equation}
\left\{
\begin{IEEEeqnarraybox}[\IEEEeqnarraystrutmode
\IEEEeqnarraystrutsizeadd{}{}][c]{rCl}
 x_0 &=& t_1\\
 x_1 &=& t_2 + t_3 + t_4\\
 x_2 &=& t_5 + t_6 + t_7 + t_8\\
 x_3 &=& t_9 + t_{10}\\
 x_4 &=& t_{11} + t_{12} + t_{13}\\
 x_5 &=& t_{14} + t_{15}\\
 x_6 &=& t_{16} + t_{17}\\
\end{IEEEeqnarraybox}
\right. \hspace{1cm} \left\{
\begin{IEEEeqnarraybox}[
\IEEEeqnarraystrutmode
\IEEEeqnarraystrutsizeadd{}
{}][c]{rCl}
 y_0 &=& t_2 + t_9 + t_{11}\\
 y_1 &=& t_5\\
 y_2 &=& t_6 + t_{14}\\
 y_3 &=& t_{12}\\
 y_4 &=& t_3 + t_{7}\\
 y_5 &=& t_{16}\\
 y_6 &=& t_1 + t_4 + t_8 + t_{10} + t_{13} + t_{15} + t_{17}\\
\end{IEEEeqnarraybox}
\right.
\label{eq:example_left_right2}
\end{equation}

 We get 626 solutions for these equations. But we observe 76 solutions lead to an invalid propagation. Hence, in order to model $\beta . S^{i}[0]$, we adopt the technique from~\cite{ElSheikhY21,zhang2019division} where the invalid propagations are eliminated by checking whether the corresponding sub-matrices of the linear transformation matrix are invertible or not. Algorithm~\ref{algo:milp_Linear_Layer} and \ref{algo:milp_FBK} elaborates the MILP model for this operation of WG-7 cipher.

\begin{algorithm}[!htb]
 \caption{MILP model for FBK operation in WG-7}
\label{algo:milp_FBK}
\begin{algorithmic}[1]

  \Function{FBK}{S, [0,11]}
  
  \State $\mathcal{M}.var \gets z_j \text{ as binary for }0\leq j \leq 6$
  \For {$i \in [0,11]$}
   \State $\mathcal{M}.var \gets s^{'}_{7i + j} \text{ , } x_{7i + j} \text{ as binary for }0\leq j \leq 6$
  \EndFor

  \For {$i \in [0,11]$}
   \State $\mathcal{M}.con \gets s_{7i+j} = s^{'}_{7i + j} + x_{7i + j} \text{ as binary for }0\leq j \leq 6$
  \EndFor
  
 \State $(\mathcal{M},Y)\gets$ {\sc LinearLayer(X)} \Comment{$X = \{x_0,x_1,\ldots,x_6\}$}

\For {$j \in [0,1,2,3,4,5,6]$}
\State $\mathcal{M}.con \gets z_j = y_j + x_{77+j}$
\EndFor
  
  \For {$j \in \{0,1,\ldots,22\}\setminus \{0,11\}$}
  \State $S^{'}[j] = S[j]$ \Comment{$S^{'}[j] = (s^{'}_{7j},s^{'}_{7j+1},s^{'}_{7j+2},s^{'}_{7j+3},s^{'}_{7j+4},s^{'}_{7j+5},s^{'}_{7j+6})$}  
  \EndFor

  \State {\Return {($\mathcal{M},S^{'},[z_0,z_1,z_2,z_3,z_4,z_5,z_6]$})}
  
  \EndFunction

\end{algorithmic}
\end{algorithm}

\noindent
\textbf{MILP model for the WG-permutation (WGP)}. We represent this function as a 7-bit S-box, where $(x_0,x_1,x_2,x_3,x_4,x_5,x_6)$
and $(y_0,y_1,y_2,y_3,$ $y_4,y_5,y_6)$ are the input and the output of WGP, respectively.
Then we use the table-aided bit-based division property on this function introduced in~\cite{DBLP:conf/crypto/BouraC16}.
After that what has been proposed in~\cite{XiangZBL16}, by using the inequality generator() function in the Sage software\footnote{http://www.sagemath.org/}, a set of 8831 inequalities returned. This makes the size of the corresponding MILP problem too large to solve. Hence, we reduce this set by the greedy algorithm in~\cite{sun2014automatic,XiangZBL16}.
We see that following 21 inequalities are sufficient as constraints to model this function.
Using these 21 inequality constraints, the MILP model for the WG-permutation
(WGP) is constructed as in Algorithm~\ref{algo:milp_wgp}. 
\begin{eqnarray*}
 1.\hspace{0.2 cm} \lefteqn{- 2x_6 - x_5 - 2x_4 - 2x_3 - 2x_1 - 2x_0 - 2y_6 + 2y_5 - y_4 + y_3 +  y_2}\\ & - 3y_1 + y_0  \geq -12\\
2.\hspace{0.2 cm}\lefteqn{- 3x_6 - 3x_5 - 2x_4 - 4x_3 - 3x_2 - 4x_1 - x_0 + y_6 - y_5 + 3y_4 + y_3 }\\ &+ 2y_2 + 2y_1 + 2y_0  \geq -17\\
3.\hspace{0.2 cm}    \lefteqn{x_6 + x_5 + x_4 + x_3 + x_2 + 25x_1 + x_0 - 5y_6 - 5y_5 - 5y_4 - 5y_3\phantom{0}- }\\ & 5y_2 - 5y_1 - 5y_0  \geq -4\\
4.\hspace{0.2 cm} \lefteqn{6x_6 - y_6 - y_5 - y_4 - y_3 - y_2 - y_1 - y_0  \geq -1}\\
5.\hspace{0.2 cm} \lefteqn{x_6 + x_5 + x_4 + x_3 + x_2 + x_1 + 29x_0 - 6y_6 - 5y_5 - 6y_4 - 5y_3 \phantom{0}- }\\ &6y_2 - 6y_1 - 6y_0  \geq -5\\
6.\hspace{0.2 cm} \lefteqn{x_6 + x_5 + 25x_4 + x_3 + x_2 + x_1 + x_0 - 5y_6 - 5y_5 - 5y_4 - 5y_3 \phantom{0}- }\\ &5y_2 - 5y_1 - 5y_0  \geq -4\\
7.\hspace{0.2 cm} \lefteqn{x_6 + x_5 + x_4 + 28x_3 + x_2 + x_1 + x_0 - 5y_6 - 6y_5 - 6y_4 - 5y_3 \phantom{0}- }\\ &5y_2 - 6y_1 - 6y_0  \geq -5\\
8.\hspace{0.2 cm} \lefteqn{10x_5 + x_2 - 3y_6 - y_5 - 2y_4 - 2y_3 - 2y_2 - y_1 - 3y_0  \geq -3}\\
9.\hspace{0.2 cm} \lefteqn{- 13x_6 - 12x_5 - 11x_4 - 9x_3 - 8x_2 - 11x_1 - 13x_0 - y_6 + 5y_5 \phantom{0}+}\\ & 2y_4 + 2y_3 - 3y_2 + y_1 + 4y_0  \geq -67\\
10.\hspace{0.2 cm} \lefteqn{- 2x_6 - x_5 - 2x_4 - 3x_3 - 2x_2 - 3x_1 - x_0 + 13y_6 + 11y_5 + 12y_4}\\ & + 12y_3 + 13y_2 + 13y_1 + 13y_0  \geq 0\\
11. \hspace{0.2cm}\lefteqn{6x_2 - y_6 - y_5 - 2y_4 - 2y_3 - y_2 - y_1 \geq -2}\\
12.\hspace{0.2 cm} \lefteqn{- 2x_5 - x_3 - x_2  - x_0 - y_6 - y_5 - y_4 - y_3 + 5y_2 - y_1 - y_0  \geq -6}\\
13.\hspace{0.2 cm} \lefteqn{- 3x_5 - 5x_4 - 4x_3 - 6x_2 - 2x_1 - 2x_0 - 4y_6 - 5y_5 + 16y_4 - y_3}\\ & - 3y_2 - 3y_1 - 3y_0  \geq -25\\
&& \phantom{00000000000000000000000000000000000000}
\end{eqnarray*}

\begin{eqnarray*}
14.\hspace{0.2 cm} \lefteqn{- 3x_6 - 4x_5 - 3x_4 + 3x_3 - 3x_2 - 3x_1 - 2x_0 + 3y_6 - 2y_5 - 5y_4}\\ & - 4y_3 - 4y_2 - 2y_1 + 9y_0  \geq -20\\
15.\hspace{0.2 cm} \lefteqn{- 2x_6 - 2x_5 - 2x_4 - 3x_3 - 2x_1 - 2x_0 + 2y_6 + y_5 - 2y_4 - y_3\phantom{0} +}\\ & 5y_2 - 3y_1 - 4y_0  \geq -15\\
16.\hspace{0.2 cm} \lefteqn{- x_6 - 3x_5 - 3x_4 - 2x_3 - 3x_1 - x_0 - y_6 + 3y_5 + 2y_4 + 2y_3\phantom{0} +}\\ & y_2 + y_1 + 3y_0  \geq -11\\
17.\hspace{0.2 cm} \lefteqn{- 3x_6 - 2x_4 - 2x_3 - x_2 - 3x_1 - 3x_0 + 4y_6 + 2y_5 + 3y_4 + 3y_3}\\ & + y_2 + 3y_1 + 4y_0  \geq -10\\
18.\hspace{0.2 cm} \lefteqn{- 2x_6 - 2x_4 - 2x_3 - 2x_2 - x_1 - x_0 + 10y_6 + 9y_5 + 9y_4 + 8y_3}\\ & + 9y_2 + 9y_1 + 10y_0  \geq 0\\
19.\hspace{0.2 cm} \lefteqn{- x_5 - x_3 - 3x_2 - x_1 + x_0 - y_5 + 6y_4 - 2y_3 - 2y_2 - 2y_1 \phantom{0}-}\\ & 2y_0  \geq -8\\
20.\hspace{0.2 cm} \lefteqn{- 6x_6 - 3x_5 - 5x_4 - 2x_3 - x_2 - 5x_1 - 6x_0 + 4y_6 + 5y_5 + 6y_4}\\ & + 6y_3 + y_2 + 5y_1 + 4y_0  \geq -21\\
 21.\hspace{0.2 cm} \lefteqn{- x_6 - 2x_4 - 2x_3 - 2x_2 - 2x_1 - x_0 + y_6 - y_5 + y_4 + 2y_3 + y_1}\\ & + y_0  \geq -9\\
 && \phantom{00000000000000000000000000000000000}
\end{eqnarray*}

\noindent
\textbf{MILP model for the Keystream Generation Operation (KSG)}. The MILP model for KSG is realized by the \textbf{COPY}, \textbf{XOR} and \textbf{AND} operations based on its ANF expression(see section~\ref{sec:Desc}).
Algorithm~\ref{algo:milp_KSG} explains the MILP variables and constraints to propagate
the division property for computing the keystream bit. Here, $[a:b]$ contains $(b-a+1)$ elements. The values lie consecutively between $a$ and $b$ such as $\{a,a+1,\ldots,b-1, b\}$. The MILP models for the \textbf{AND} and \textbf{XOR} are described in Algorithms~\ref{algo:milp_AND} and \ref{algo:milp_XOR}, respectively. 
The overall MILP model for the WG-7 whose initialization is
reduced to \textit{R} rounds is given as function \textbf{WG7Eval} in Algorithm~\ref{algo:milp_WG7}. The required number(s) of MILP variables and constraints for \textbf{WG7Eval} are 
$(73R + 532)$ and $(78R + 584)$, respectively.

\subsection{\bf Evaluate Secret Variables Involved in the Superpoly}\label{subsec:sec_key_involve}%
We start the evaluation by preparing a cube $C_{I}(IV)$ taking all possible combinations of
$\{v_{i_1},v_{i_2},\ldots,v_{i_{|I|}}\}$.  Then, we extract the involved secret variables
$J = \{k_{j_1},k_{j_2},\ldots,k_{j_{|J|}}\}$ in the preferable superpoly using
the technique proposed in[Algorithm 1,~\cite{TodoIHM17}]. It computes all the $R$-round division trails taking the initial division
property as $v_i = 1$ for $i\in I$, and $v_i = 0$ for $i\in\{0,1,\ldots,m-1\}\setminus I$. Table~\ref{table:DT_involved_key_set2_1}
summarize all the secret bits involved in the preferable superpoly.  Table~\ref{table:DT_involved_key_set2_1} is
built for 14 to 20 rounds of the initialization phase based on our chosen cubes
of Table~\ref{table:DT_involved_key_set2_1}.

\noindent
\textbf{Searching Cubes.} We choose the cube $I$ such that the value of $2^{I + J}$ becomes as small as possible.
The cubes that we choose for this attack satisfy the above condition, and are
shown in Table~\ref{table:DT_involved_key_set2_1} as $I_1,I_2,\ldots,I_{8}$.
A total of $\binom{81}{|I|}$ cubes of size $|I|$ are possible. It is computationally infeasible to choose so many cubes. We do not have the reasonable evidence that our choice of cube indices are appropriate. But we have experimented a lot cubes randomly. Based on our experiments,
the cubes of Table~\ref{table:DT_involved_key_set2_1} appear
to be the best for this cipher. How to choose appropriate cubes is left as an open question.

\noindent
\textbf{Extract a balanced superpoly.} We choose the constant part of the IV randomly, and recover the superpoly $p(J,\bar{v})$, for $\bar{v}=\{v_0,v_1,\ldots,v_{m-1}\}\setminus I$ by trying out a total of
$2^{7}\times 2^{62}$ possible combinations (see Table~\ref{table:DT_involved_key_set2_1}) for the reduced 20-round of the cipher. Let $\hat{J}$ be the set of $2^{62}$ possible values of $J$.
In the offline phase, we compute the values of $p(J,\bar{v})$, and store them in
a table $T_1$ indexed by $\hat{J}$, and then evaluate the ANF accordingly. If $p(J,\bar{v})$
becomes constant, we pick another random $IV$, and repeat the above process until we find
an appropriate one such that $p(J,\bar{v})$ is not constant. 

\begin{minipage}{0.44\textwidth}

\begin{algorithm}[H]
 \caption{MILP model for AND operation in WG-7}
\label{algo:milp_AND}
\begin{algorithmic}[1]

  \Function{AND}{S, \textit{I}}
  \For {$i \in I$}
   \State $\mathcal{M}.var \gets s^{'}_{i}\text{ , } x_i \text{ as binary}$
  \EndFor
  \State $\mathcal{M}.var \gets y \text{ as binary}$
  
  \For {$i \in I$}
   \State $\mathcal{M}.con \gets s_i = s^{'}_{i} + x_i $
  \EndFor
  
  \For {$i \in I$}
   \State $\mathcal{M}.con \gets y \geq x_i $
  \EndFor
  
  \For {$k \in (0,1,\ldots,160) \setminus I$}
  \State $s^{'}_k = s_k$  
  \EndFor
  
  \State {\Return {($\mathcal{M},S^{'},y$})}
  
  \EndFunction

  \end{algorithmic}
\end{algorithm}

\end{minipage}
\hfill
\begin{minipage}{0.45\textwidth}
\begin{algorithm}[H]
 \caption{MILP model for XOR operation in WG-7}
\label{algo:milp_XOR}
\begin{algorithmic}[1]

  \Function{XOR}{S, \textit{I}}
  
  \For {$i \in I$}
   \State $\mathcal{M}.var \gets s^{'}_{i}\text{ , } x_i \text{ as binary}$
  \EndFor
  \State $\mathcal{M}.var \gets y \text{ as binary}$
  
  \For {$i \in I$}
   \State $\mathcal{M}.con \gets s_i = s^{'}_{i} + x_i $
  \EndFor
  
  \State $\text{temp } = 0$
  
  \For {$i \in I$}
   \State $\text{temp = temp } + x_i $
  \EndFor
  
  \State $\mathcal{M}.con \gets y = temp$
  
  \For {$k \in (0,1,\ldots,160) \setminus I$}
  \State $s^{'}_k = s_k$  
  \EndFor
  \State {\Return {($\mathcal{M},S^{'},y$})}
  \EndFunction

\end{algorithmic}
\end{algorithm}
\end{minipage}

\begin{algorithm}[H]
 \caption{MILP model for the Initialization Round of WG-7 Stream Cipher}
\label{algo:milp_WG7}
\begin{algorithmic}[1]

  \Function{WG7Eval}{R}
  \State \text{Prepare an empty MILP model $\mathcal{M}$}
  \State $\mathcal{M}.var \gets S^0[i] \text{ for } 0\leq i \leq 22$ \Comment{$S^0[i] = (s_{7i},s_{7i+1},s_{7i+2},\ldots,s_{7i+6})$}
  \For {$r \in \{1,2,\ldots, R\}$}
  \State $(\mathcal{M},S^{'}, a)\gets WGP(S^{r-1})$
  \State $(\mathcal{M},S^{''}, b)\gets FBK(S^{'},[0,11])$
  \For {$i = 0 \text{ to } 21$}
  \State $S^r[i] = S^{''}[i+1]$
  \EndFor
  \State $\mathcal{M}.con \gets S^{''}[0] = 0$
  \State $\mathcal{M}.var \gets S^r[22] \text{ as binary}$
  \State $\mathcal{M}.con \gets S^r[22] = a + b$
  \EndFor
  
  \State $(\mathcal{M},S^{'''},z)\gets KSG(S^{R})$
  \For {$i= 0 \text{ to } 22$}
  \State $\mathcal{M}.con \gets S^{'''}[i] = 0$
  \EndFor
  \State $\mathcal{M}.con \gets z = 1$
  
  \EndFunction

\end{algorithmic}
\end{algorithm}

\begin{algorithm}[!h]
 \caption{MILP model for the WG-permutation in WG-7}
\label{algo:milp_wgp}
\begin{algorithmic}[1]

  \Function{WGP}{S}
  \State $\mathcal{M}.var \gets s^{'}_{154 + i}\text{ , } x_{i} \text{ , } y_{i} \text{ as binary for } 0\leq i\leq 6$
  \State $\mathcal{M}.con \gets s_{154 + i} = s^{'}_{154 + i} + x_i \text{ for }0\leq i\leq 6$
  \State Add constraints to $\mathcal{M}$ based on the reduced set of inequalities

  \For {$j = 0 \text{ to } 21$}
  \State $S^{'}[j] = S[j]$ \Comment{$S^{'}[j] = (s^{'}_{7j},s^{'}_{7j+1},s^{'}_{7j+2},s^{'}_{7j+3},s^{'}_{7j+4},s^{'}_{7j+5},s^{'}_{7j+6})$}  
  \EndFor
  
  \State {\Return {($\mathcal{M},S^{'},[y_0,y_1,y_2,y_3,y_4,y_5,y_6]$})}
  
  \EndFunction

\end{algorithmic}
\end{algorithm}

To sum up, we compute a table $T_1$ of size $\hat{J}$ by computing $2^{7+62}$ operations.
The attack is possible if the attacker can find the appropriate IVs easily.
We assume that we can recover the balanced superpoly in only one trial for each of the cubes
in Table~\ref{table:DT_involved_key_set2_1}. Indeed, since each of these cubes has size 7,
there exist $81-7 = 74$ bits to set the constant part of the IV. Therefore, we anticipate
that Assumption~1 (mentioned in~\cite{TodoIHM17}) holds with high probability, and derive the complexity figures
accordingly.

\subsection{\bf Key Recovery for $20$-Round Initialization Phase}\label{subsec:eval_time_comp}%
We use the balanced superpolys for the cubes $I_1, I_2, \ldots,I_{8}$.
The online phase consists of the following operations for each $i \in \{1,2,\ldots,8\}$.

\begin{itemize}
 \item Query the encryption oracle with $C_{I_i}$ and compute $S=\bigoplus_{C_{I_i}}f(k,v)$
 \item Compare $S$ with each entry of $T_1$. 
The values of $\bar k = \{k_{j_1},k_{j_2},\ldots,k_{j_{62}}\}$ for which $S$ does not match
$T_1$ are discarded. Since the superpoly is balanced, we have
$p(\{k_{j_1},k_{j_2},\ldots,k_{j_{62}}\}\},\bar{v}) = 0$ for $2^{61}$ values of
$\bar k$, and
$p(\{ k_{j_1},k_{j_2},\ldots,k_{j_{62}}\},\bar{v}) = 1$ for the remaining $2^{61}$
values of $\bar k$. Therefore, we can recover one bit of
information in the secret variables.
\end{itemize}
We can recover one bit of secret information for each cube only in one trial.
Since we work with eight cubes, we recover eight secret variables.
The remaining secret bits ($80-8 = 72$ of them) are recovered by guessing
involving a brute-force complexity of $2^{72}$. The total time complexity for
the attack is therefore $8\times 2^{69} + 2^{72} = 2^{73}$.
The data complexity for the total computation is $8\times 2^{7} = 2^{10}$.

\begin{table}[!htb]
\begin{center}
\caption{Involved Key bits in the Superpoly for the Cube $C_{\{I_1, I_2,\ldots, I_8\}}$}
\label{table:DT_involved_key_set2_1}
\resizebox{\columnwidth}{!}{
\begin{tabular}{|cc|c|c|c|c|} 
\hline
 \thead{Cube\\ Sequence} & \thead{Cube\\ Indices} &\text{\#Rounds}&  Involved Secret Key Variables ($J$)& $|J|$ & Time\\
\hline
$I_1$&0,36,37,73,74,75,76&14&  0, 1, 2,\ldots, 5, {\bf 6,39},40,\ldots,47,{\bf 48,77},78,79& 20 & $2^{20 + 7}$\\
\cline{3-6}
$I_2$&35,37,73,74,75,76,80&15&  0, 1, 2,\ldots, 9, {\bf 10,39},40,\ldots,51,{\bf 52,77},78,79& 28 & $2^{28 + 7}$\\
\cline{3-6}
$I_3$&36,37,73,74,75,76,78&16&  0, 1, 2,\ldots, 12, {\bf 13,39},40,\ldots,54,{\bf 55,77},78,79& 34 & $2^{34 + 7}$\\
\cline{3-6}
$I_4$&35,37,73,74,75,76,77&17& 0, 1, 2,\ldots, 16, {\bf 17,39},40,\ldots,58,{\bf 59,77},78,79& 42 & $2^{42 + 7}$\\
\cline{3-6}
$I_5$&36,37,45,73,74,75,76&18& 0, 1, 2,\ldots, 19, {\bf 20,39},40,\ldots,61,{\bf 62,77},78,79& 48 & $2^{48 + 7}$\\
\cline{3-6}
$I_6$&35,37,73,74,75,76,79&19& 0, 1, 2,\ldots, 23, {\bf 24,39},40,\ldots,65,{\bf 66,77},78,79& 56 & $2^{56 + 7}$\\
\cline{3-6}
$I_7$&35,37,38,73,74,75,76&20& 0, 1, 2,\ldots, 26, {\bf 27,39},40,\ldots,68,{\bf 69,77},78,79& 62 & $2^{62 + 7}$\\
%\cline{2-5}
$I_8$&35,37,39,73,74,75,76 &&&&\\
\hline

\end{tabular}
}
\end{center}
\end{table}

\section{Conclusion}\label{sec:conclusion}%
This paper investigates the security of the nonlinear feedback-based initialization phase of the lightweight stream cipher WG-7~\cite{LuoCGL10}.
We mount a division-property-based cube attack by considering the structure of the cipher
as a non-blackbox polynomial. Our attack proposes a MILP model for the initialization
phase and the keystream-generation phase of the cipher. We work out the details of the
model as specific to the WG-7 cipher. To the best of our knowledge, the only two published attacks against this cipher are both algebraic attack that can work only
when no non-linear filter function present in the feedback path of the cipher. Our attack works both in the presence and in
the absence of non-linear filter functions, and reduces the data complexity over
those algebraic attacks.

\bibliographystyle{splncs04}
%\bibliography{WG7}

\end{document}